\documentclass[prl,superscriptaddress,twocolumn, showpacs,preprintnumbers]{revtex4}

\usepackage{natbib}

\usepackage{amsmath}
\usepackage{amssymb}
\usepackage{amsthm}
\usepackage{amscd}

\usepackage{graphicx}
\usepackage{subfigure}
\usepackage{psfrag}

\usepackage{bm}
\usepackage{cancel}

\usepackage[mathscr]{eucal}

\graphicspath{{/home/participants/heinzle/figures/}}

\newcommand{\nextparagraph}{\parskip2ex}

\newcommand{\textfrac}[2]{{\textstyle \frac{#1}{#2}}}

\newcommand{\weg}{\:\,}

\DeclareMathOperator{\tr}{tr}

\DeclareMathOperator{\diag}{diag}

\theoremstyle{plain}

\theoremstyle{remark}
\newtheorem*{Remark}{Remark}

\begin{document}

\title{On closed cosmological models that satisfy the strong energy condition 
but do not recollapse}

\author{Simone Calogero}
\email{calogero@ugr.es}
\affiliation{Departamento de Matem\'atica Aplicada,
Facultad de Ciencias, Universidad de Granada
18071 Granada, Spain}

\author{J.~Mark Heinzle}
\email{Mark.Heinzle@univie.ac.at}
\affiliation{Gravitational Physics, Faculty of Physics, University of Vienna,
1090 Vienna, Austria}

\begin{abstract}
We show the existence of a rather general class of closed cosmological models 
of Bianchi type~IX 
that do not exhibit recollapse but expand for all times. This is despite
the fact that these models satisfy the strong energy condition by a wide margin.
\end{abstract}

\pacs{04.20.-q, 04.20.Cv, 98.80.Jk}
 
\maketitle


In relativistic cosmology, the 
trichotomy of the Fried\-mann-Robertson-Walker (FRW) models
is of prime importance.
The spatial geometry 
determines the evolution of the cosmological model:
In the hyper\-bo\-loidal ($k = {-1}$) or the 
spatially flat ($k = 0$) case, the universe exhibits an initial singularity
(`big bang'), and from that `moment' on, the universe is forever expanding.
In the case of a closed cosmological model ($k={+1}$)
we observe a fundamentally different behavior:
``[\ldots] the dynamical equations of general relativity
show that the spatially closed 3-sphere universe will exist
for only a finite span of time.
[\ldots] at a finite time after the big bang, the universe
will achieve a maximum size [\ldots], and then will begin to
recontract. [\ldots] a finite time after recontraction begins, a `big crunch'
will occur.'' The quotation is taken from~\cite{Wald:1984}.
The recollapse of closed FRW cosmologies holds because
the \textit{strong energy condition} is imposed on the matter 
(which is assumed to be a perfect fluid), i.e., $\rho + 3 p \geq  0$,
where $\rho$ is the energy density and $p$ the pressure.


On this basis it would be tempting to view the 
recollapse of the spatially closed FRW cosmologies 
satisfying the strong energy condition
as a paradigm for models with less symmetries.
That this belief is erroneous
has been demonstrated in~\cite{Barrow/Galloway/Tipler:1986}.
At least in the case of cosmological models of Bianchi type~IX, 
into which the closed FRW models
are naturally embedded, there exists a rigorous result by 
Lin and Wald~\cite{Lin/Wald:1989}:
Assuming the dominant energy condition, i.e., $|p_i| \leq \rho$ 
for the anisotropic pressures $p_1$, $p_2$, $p_3$,
and non-negative average principal pressure $p$, i.e., 
$3 p = (p_1+p_2+p_3) \geq 0$, then
the `closed-universe-recollapse conjecture'~\cite{Barrow/Galloway/Tipler:1986} holds.
In the locally rotationally symmetric (LRS) case \textit{with isotropic matter}, 
it is sufficient to require that $\rho + 3 p \geq \epsilon \rho$ 
for an arbitrarily small $\epsilon>0$, see~\cite{Heinzle/Rohr/Uggla:2005}.
(Note that if $p/\rho \rightarrow -1/3$, there exist models that expand forever
approaching the Einstein static universe in the limit.)

In this letter, 
we approach the problem from a different direction
by investigating cases where `closed-universe-recollapse'
does not hold.
We prove that there exist 
cosmological models (of Bianchi type~IX) 
satisfying the strong energy condition
that do not recollapse but expand forever.
Two points are important to emphasize:
(i) For these models 
the strong energy condition is satisfied `by a wide margin'.
The assumption we make is that $w = p/\rho$ is a constant, i.e.,
$w = \mathrm{const} > -1/3$; however, $w < (1-\sqrt{3})/3 \approx -0.244$,
hence the average pressure is not positive.
(ii) We prove the existence of a \textit{typical} 
class of models that expand forever (where `typical' is 
understood in the sense of an open set of initial data of the Einstein
equations).
An interesting observation is that these cosmological models
exhibit partial (i.e., directional) accelerated expansion
for late times.

As a matter of course, we 
do not propose the cosmological models we analyze
as actual models of the universe.
However, we want to emphasize that the matter model we consider
is not `exotic'~\cite{Vollick:1997}, i.e., it satisfies all the standard 
energy conditions (weak, strong and dominant), as opposed to certain `exotic' 
matter models in cosmology (e.g., phantom fields and 
dark energy, see for instance ~\cite{Amendola:2004, Scherrer/Sen:2008};
the breaking of the energy conditions 
stems from the aim to account for the accelerated expansion of the universe).
The properties of the matter source we consider in this paper resemble  
those of collisionless matter, elastic matter,
and magnetic fields as regards their fundamental aspects~\cite{Calogero/Heinzle:2009, Calogero/Heinzle:2009a}.
Classes of models encompassing these important examples (or a subset thereof)
have been the basis of previous work, see, e.g.,~\cite{Barrow}.
Another explicit example, an example that satisfies our concrete assumptions,
is the anisotropic fluid model~\cite{Ellis}.
For the models we consider the anisotropic pressures (parallel and perpendicular pressure)
are required to satisfy certain bounds (that are compatible with
the energy conditions). In particular, the isotropic pressure 
(average pressure) depends linearly on the energy density,
as is usual for perfect fluids, where the proportionality
constant is strictly larger than $-\textfrac{1}{3}$.
Note, however, that the approach we take does not require
to specify a concrete matter model as long as the basic assumptions
of~\cite{Calogero/Heinzle:2009, Calogero/Heinzle:2009a} and the
necessary bounds on the anisotropic pressures are satisfied.
Finally, note that the restriction to matter sources 
of the general type of~\cite{Calogero/Heinzle:2009a, Calogero/Heinzle:2009} might
yield a special case of the `closed-universe-recollapse conjecture' 
and, in view of the results of this paper, 
lead to specific bounds on the matter quantities.

Let us briefly comment on the approach we take.
We use the dynamical systems approach to spatially
homogeneous cosmologies~\cite{Wainwright/Ellis:1997}.
However, as we will see, it is essential to avoid
the standard Hubble-normalized variables---the 
results we present here are rather elusive in that
approach.


\nextparagraph
\paragraph{Equations.}


For models of Bianchi type~IX the metric can be written as
\begin{equation}\label{metric}
d s^2 = -d t^2 + g_{i j}(t) \:\hat{\omega}^i\, \hat{\omega}^j\:,
\end{equation}
where $\{\hat{\omega}^1,\hat{\omega}^2,\hat{\omega}^3\}$ is a symmetry-adapted coframe
satisfying $d\hat{\omega}^1 = - \hat{\omega}^2 \wedge \hat{\omega}^3$ (and cyclic permutations).
The Einstein equations comprise evolution equations for the metric,
$\partial_t g_{i j} = -2 g_{i l} k^l_{\weg j}$, and 
for the extrinsic curvature $k^i_{\weg j}$, see, e.g.,~\cite{Wainwright/Ellis:1997}.
The Gauss constraint reads
${}^3\!R +(\tr k)^2 - k^i_{\weg j} k^j_{\weg i} = 2 \rho\,$,
where $\rho = -T^0_{\weg 0}$ is the energy density associated with the energy-momentum tensor $T^\mu_{\weg \nu}$
and ${}^3\!R$ the scalar three-curvature.
In the vacuum case or orthogonal perfect fluid case the metric is 
diagonal, i.e., $g_{i j}(t) = \diag\big( g_{11}(t), g_{22}(t), g_{33}(t)\big)$;
in the locally rotationally symmetric (LRS) case we have $g_{22}(t) \equiv g_{33}(t)$.
We restrict ourselves to diagonal metrics even if the matter source
is anisotropic.

In the diagonal case, an isotropic matter source is
characterized by an energy-momentum tensor 
with $T^i_{\weg j} = \diag\big( p_1, p_2, p_3\big)$.
The `isotropic pressure' $p$ is the average of the anisotropic pressures $p_1$, $p_2$, $p_3$,
i.e., $\tr T = 3 p$.
Define $w$ and $w_i$, $i=1,2,3$,
according to
\begin{equation}
p = w \rho\:,\qquad p_i = w_i \rho \:; 
\end{equation}
obviously, $w_1 + w_2 + w_3 = 3 w$. 
Matter that is consistent with LRS symmetry satisfies $w_2 = w_3$.
For perfect fluids, $w_1 = w_2 = w_3 = w$, where $w$ is typically
assumed to be a constant. 
In this paper we consider anisotropic matter that 
generalizes perfect fluid matter:
We assume that the energy density and the isotropic pressure
satisfy a linear equation of state, $w=\mathrm{const}$,
where the strong energy condition is supposed to hold, i.e.,
$w > -1/3$.
The rescaled anisotropic pressures, $w_1$, $w_2$, $w_3$,
are assumed to be functions of the metric via $(s_1,s_2,s_3)$, 
where
\begin{equation}\label{skdef}
s_k = g^{kk} \big(g^{11} + g^{22} + g^{33}\big)^{-1} 
\quad(\text{no sum over $k$})\:;
\end{equation}
obviously, $s_1 + s_2 + s_3 =1$.
The functions 
\begin{equation}
w_k = w_k(s_1,s_2,s_3)
\end{equation}
are such that there exists
an isotropic state 
of the matter where $w_1 = w_2 = w_3 = w$, and remain bounded (and take
limits) under extreme conditions (when one or more
of the $s_i$ are zero). In particular, there exists a 
constant $v_-$ such that 
$w_1(0,s_2,s_3) = w_2(s_1,0,s_3) = w_3(s_1,s_2,0) = v_-$.
There exist excellent
examples for matter models of this type, e.g.,
collisionless matter, elastic matter, or magnetic fields;
for a detailed discussion we refer to~\cite{Calogero/Heinzle:2009}.

Let us define the Hubble scalar $H$ as $H= -\tr k/3$ and
the shear tensor $(\sigma_1,\sigma_2,\sigma_3)$ as the traceless part of the extrinsic curvature, i.e.,
$k^i_{\weg i} = -H - \sigma_i$ (no sum over $i$); 
$\sigma_1 + \sigma_2 + \sigma_3 = 0$.
Furthermore, we introduce the `densitized metric' $(n_1,n_2,n_3)$ by
$n_k = g_{k k} (\det g)^{-1/2}$;
note that $n_k > 0$. Then the Einstein equations can be expressed as 
evolution equations for $H$, $(\sigma_1,\sigma_2,\sigma_3)$, and $(n_1,n_2,n_3)$
plus one constraint, which can be used to express $\rho$
in terms of the other variables; 
this leads to the fact that the matter enters the equations
only via $w$ and $w_1$, $w_2$, $w_3$.

In the LRS case, which we will focus on henceforth,
there exists a plane of rotational symmetry, which
we choose to be spanned by the second and the third frame
vector. Accordingly, 
$n_2 = n_3$ and $\sigma_2 = \sigma_3$ 
as well as $s_2 = s_3$; consistently, the matter 
satisfies $w_2 = w_3$.
Let $\sigma := \sigma_2$ and let $s := s_2$; then 
\begin{equation}\label{sigs}
(\sigma_1, \sigma_2, \sigma_3) = ({-2}\sigma, \sigma, \sigma),\;
(s_1,s_2,s_3) = (1-2s,s,s).
\end{equation}
Eq.~\eqref{skdef} implies 
$s = (2 + n_2/n_1)^{-1}$,
so that 
$s\in(0,1/2)$.
Finally we abbreviate the rescaled anisotropic pressure in the
plane of rotational symmetry by $u$; more specifically,
\begin{equation}\label{uofsdef}
u(s) := w_2(1-2s,s,s)\:.
\end{equation}
The Einstein equations can then be expressed in 
the variables $H$, $\sigma$, $n_1$, and $n_2$,
where $u(s)$ appears in these equations.

In the dynamical system approach to cosmology the Einstein equations are 
expressed in terms of normalized variables. 
We define the `dominant variable' $D$, see, e.g.,~\cite{Heinzle/Rohr/Uggla:2005}, by
\begin{subequations}\label{domvars}
\begin{equation}\label{Ddef}
D = \sqrt{H^2 + \frac{n_1 n_2}{3}}\:,
\end{equation}
and we introduce normalized variables according to
\begin{equation}\label{normvars}
\Sigma_D = \frac{\sigma}{D} \:,\quad
r = \frac{n_1}{D}\,\sqrt{\frac{n_1^2}{D^2} + \frac{n_2^2}{9 D^2}}\:;
\end{equation}
in addition we use the variable
\begin{equation}\label{sinn1n2}
s = \Big(2 + \frac{n_2}{n_1}\Big)^{-1} \:.
\end{equation}
\end{subequations}
Further, we define a normalized energy density 
$\Omega_D = \rho/(3 D^2)$,
and we replace the cosmological time $t$ by a rescaled time variable $\tau$ through
\begin{equation}\label{newtime}
\frac{d}{d\tau} = \frac{1}{D} \,\frac{d}{d t}\:.
\end{equation}
Henceforth, a prime denotes differentiation w.r.t.\ $\tau$.

\begin{Remark}
The evolution equation for $H$ is 
\begin{equation}\label{Heq}
H^\prime = -\frac{1}{D} \,\big( H^2 + q_D D^2 \big) \:,
\end{equation}
where $q_D$ is given by
$q_D = 2 \Sigma_D^2 + (1/2)( 1 + 3 w) \Omega_D$.
This leads to an important remark:
Equation~\eqref{Heq} implies that $H$ is decreasing if $H = 0$, i.e., 
$H^\prime |_{H = 0} = -q_D D^{-1}< 0$; therefore, a cosmological model with
$H(\tau_0) > 0$ at some time $\tau_0$ satisfies $H(\tau) > 0$ $\forall \tau \leq \tau_0$.
Consequently, by proving the existence of models that satisfy
$H(\tau) > 0$ for all sufficiently large $\tau$, we prove the existence of models
with $H>0$ and thus positive expansion for all times.
\end{Remark}

It is not difficult to prove that
the transformation~\eqref{domvars} between
the `metric variables' $H$, $\sigma$, $n_1$, $n_2$
and the dynamical systems variables $D$, $\Sigma_D$, $r$, $s$ is
one-to-one on the set $H > 0$.
(This is sufficient for our purposes, see the previous remark.)

Expressed in the variables $D$, $\Sigma_D$, $r$, $s$, 
the Einstein evolution equations split into a decoupled
equation for $D$,
\begin{equation}\label{Ddecoupled}
D^\prime = -D \Big( H_D (1+ q_D) + \Sigma_D (1 - H_D^2)\Big)\:,
\end{equation}
and a system of coupled equations for the normalized variables~\eqref{normvars} and~\eqref{sinn1n2},
\begin{subequations}\label{dynsys}
\begin{align}
r^\prime & = r \left( 2 H_D (q_D - H_D \Sigma_D) - \frac{54 \Sigma_D s^2}{1-4 s + 13 s^2} \right) \:,\\[0.5ex]
\nonumber
\Sigma_D^\prime & = -(2- q_D) H_D\Sigma_D - (1-H_D^2) (1-\Sigma_D^2) \: + \\
\label{SigDeq}
& \qquad\qquad\qquad + \textfrac{1}{3} N_{1D}^2 + 3\Omega_D \big(u(s) -w\big) \:,\\[0.5ex]
s^\prime & = -12 s \big( \textfrac{1}{2} - s \big) \Sigma_D \:,
\end{align}
\end{subequations}
where $q_D =  2 \Sigma_D^2 + \textfrac{1}{2} ( 1 + 3 w) \Omega_D$, 
and $H_D = H_D(r,s)$ and $N_{1D} = N_{1D}(r,s)$ 
are functions of $r$ and $s$, see~\eqref{HDN1Dinrs}.
The Gauss constraint reads
\begin{equation}\label{gcon}
\Sigma_D^2 + \textfrac{1}{12} N_{1D}^2(r,s) + \Omega_D  = 1\:;
\end{equation}
it is used to solve for $\Omega_D$.
The functions $H_D(r,s)$ and $N_{1D}(r,s)$ are
\begin{subequations}\label{HDN1Dinrs}
\begin{align}
\label{HDinrs}
H_D & := \frac{H}{D} = H_D(r,s) = \sqrt{1 - r \frac{1-2s}{\sqrt{1-4s+13 s^2}}} \:\,,\\
\label{N1Dinrs}
N_{1D} & := \frac{n_1}{D} = N_{1D}(r,s) = \sqrt{\frac{3 r s}{\sqrt{1-4s +13 s^2}}}\:\,;
\end{align}
\end{subequations}
in particular,
$H_D(r,s)$ and $N_{1D}(r,s)$ are
well-defined and regular  
on the preimage of the set $\mathbb{R}^+ \times \mathbb{R}^+$.
(We refrain from going into details in this paper, since
we merely use that~\eqref{HDN1Dinrs} is well-behaved for sufficiently small $r$;
however, we may refer to~\cite{Calogero/Heinzle:2009a}.)

\begin{Remark}
More common than~\eqref{domvars} is
the Hubble-nor\-ma\-lized approach, see, e.g.,\cite{Wainwright/Ellis:1997}, 
where the Hubble scalar $H$
is employed instead of $D$ to construct scale-invariant variables 
and a simpler set of normalized variables is used instead of~\eqref{normvars} and~\eqref{sinn1n2}.
Although the resulting equations are simpler than~\eqref{dynsys}, 
we do not have a choice;
we will see that the Hubble-normalized approach
necessarily fails to uncover our results.
\end{Remark}

The dynamical system~\eqref{dynsys} completely describes
the dynamics of locally rotationally symmetric cosmological models of
Bianchi type~IX in their expanding phase ($H > 0$). In other words,
each solution of~\eqref{dynsys} yields an LRS Bianchi type~IX model
in its expanding phase, and conversely, the expanding phase of 
each model is represented by a solution of~\eqref{dynsys}.
The main advantage of the system~\eqref{dynsys} over other representations
of the Einstein equations lies in its extendibitly: The system~\eqref{dynsys}
possesses a regular extension to its boundaries $\Sigma_D = \pm 1$,
$s=0$, $s=\textfrac{1}{2}$, and $r=0$.

\nextparagraph
\paragraph{Anisotropic matter.}


The function $u(s)$ 
in~\eqref{SigDeq}
encodes the
properties of the matter model. For isotropic matter we have $u(s) \equiv w$;
for anisotropic matter, $u(s)$
represents the rescaled anisotropic pressure
in the plane of local rotational symmetry, cf.~\eqref{uofsdef};
recall that $s\in [0,1/2]$.
%
From $w_2(s_1,0,s_3) = v_-$
we obtain $u(0) = v_-$, cf.~\eqref{sigs} and~\eqref{uofsdef}. 
The value of $u$ at $s=1/2$ is not independent; to see this 
recall first that 
$w_1 + 2 w_2 = 3 w$;
now, $s = 1/2$ corresponds to $(s_1,s_2,s_3) = (0,1/2,1/2)$, so
$w_1(0,1/2,1/2) + 2 w_2(0,1/2,1/2) = 3 w$; hence, $w_1(0,s_2,s_3) = v_-$ results in 
$v_- + 2 u(1/2) = 3 w$.
Summarizing,
\begin{equation}\label{uvals}
u(0)  = v_- \:,\qquad
u(1/2) =  w + (1/2) \big(w-v_-\big)\:. 
\end{equation}
For reasonable matter models like collisionless matter, elastic matter, or magnetic fields, 
$u(s)$ is a monotone function on $[0,1/2]$
interpolating between the values~\eqref{uvals}; 
we refer to~\cite{Calogero/Heinzle:2009} for examples.
It is often beneficial to use the anisotropy parameter $\beta$ instead
of the constant $v_-$; it is defined as
\begin{equation}\label{betadef}
\beta = \frac{2 ( w- v_-)}{1-w}\:.
\end{equation}

\begin{figure}[Ht]
\begin{center}
\includegraphics[width=0.45\textwidth]{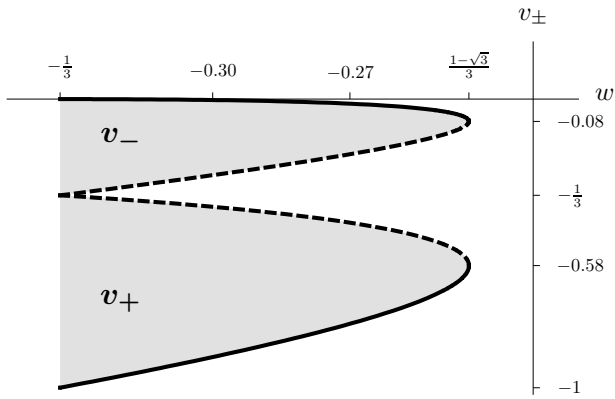}
\caption{The admissible values of the rescaled isotropic pressure $w$ and $v_-$, $v_+ = 3 w -2 v_-$, which 
represent (the extremes of) the rescaled anisotropic pressures in the 
plane of symmetry and orthogonal to it, respectively. Both the strong and the 
dominant energy condition are satisfied. $(1-\sqrt{3})/3 \approx -0.24$.}
\label{vpmminmax}
\end{center}
\end{figure}

Finally, let us state the assumptions on the matter model we consider in this paper.
We assume that
\begin{equation}\label{wdomain}
-\frac{1}{3} \,<\, w\,<\, \frac{1-\sqrt{3}}{3} \approx -0.244 \:,
\end{equation}
hence the \textit{strong energy condition is satisfied}. 
Furthermore, 
$v_-$
is assumed to satisfy 
\begin{equation}\label{vmdomain}
\begin{split}
& \textfrac{1}{6} \left( 1 + 6 w - \sqrt{-3+(1-3 w)^2} \right) < v_- \\
& \qquad\qquad v_- < \textfrac{1}{6} \left( 1 + 6 w + \sqrt{-3+(1-3 w)^2} \right).
\end{split}
\end{equation}
The admissible domain of the parameters $w$ and $v_-$ is depicted in Fig.~\ref{vpmminmax}.
Clearly, the dominant energy condition is satisfied since the rescaled
anisotropic pressure $v_-$ (which is the pressure in the plane
of symmetry) and its counterpart $v_+ = 3 w- 2 v_-$ (which is
the pressure in the orthogonal direction) satisfy $|v_\pm| < 1$.

\nextparagraph
\paragraph{Results.}


Since the dynamical system~\eqref{dynsys} extends regularly to $r=0$ 
it is suggestive to analyze the system induced on that surface.
From~\eqref{HDN1Dinrs} we obtain $H_D|_{r=0} = 1$, $N_{1D}^2|_{r=0} =0$,
so that~\eqref{gcon} becomes 
$\Omega_D = 1 -\Sigma_D^2$.
This in turn implies $q_D = \textfrac{1}{2} (1 + 3 w) + \textfrac{3}{2} (1 - w) \Sigma_D^2$.
Insertion into~\eqref{dynsys} yields
\begin{subequations}\label{dynsysonr0}
\begin{align}
\Sigma_D^\prime & = -3 (1-\Sigma_D^2) \Big( \textfrac{1}{2} (1 -w) \Sigma_D -  \big(u(s)-w \big)\Big) \:,\\
s^\prime  & = -12 \Sigma_D \,s \left(\textfrac{1}{2} - s \right)\:.
\end{align}
\end{subequations}
The state space for this two-dimensional dynamical system is 
$[-1,1] \times [0,1/2]$.
The dynamical systems analysis is straightforward.
We use that $u(s)$ is a function such that $u(0) = v_-$ 
and $u(1/2) = w + (w - v_-)/2$; 
here, $w$ and $v_-$ are assumed to satisfy~\eqref{wdomain} and~\eqref{vmdomain},
respectively. 

We focus our attention on the fixed point $\mathrm{R}$ of~\eqref{dynsysonr0}, 
which is given by
\begin{equation}
\mathrm{R}: r=0, \: s = 0, \:\Sigma_D = -\beta \:. 
\end{equation}
It is straightforward to prove that $\mathrm{R}$ is a sink for the
flow of the system~\eqref{dynsysonr0}, because
\begin{subequations}
\begin{align}
\label{sinkprop}
s^{-1} s^\prime \big|_{\mathrm{R}} & = 6 \beta \,< \,0 \:,\\[0.5ex]
\nonumber
(\Sigma_D + \beta)^{-1} (\Sigma_D + \beta)^\prime  \big|_{\mathrm{R}} & 
= -\textfrac{3}{2} ( 1- \beta^2) (1 - w) \,< \,0\:,
\end{align}%
\addtocounter{equation}{1}%
hence the eigenvalues of the linearization of~\eqref{dynsysonr0}
at $\mathrm{R}$ are negative.



The crucial property of the fixed point $\mathrm{R}$ is revealed by
considering the full system~\eqref{dynsys}:
$\mathrm{R}$ is a sink 
not only on the boundary $r=0$, but also for the full system~\eqref{dynsys}.
To see this we simply compute
\begin{equation}\label{r-1r'}
r^{-1} r^\prime \big|_{\mathrm{R}} = 2 \Big( \textfrac{3}{2} (1- w) \beta^2 + \textfrac{1}{2} ( 1 + 3 w) + \beta \Big) < 0
\end{equation}
\end{subequations}
and use~\eqref{vmdomain} to establish that the r.h.\ side is negative. 

Since the fixed point $\mathrm{R}$ is a sink for the flow of
the system~\eqref{dynsys}, i.e., for 
LRS Bianchi type~IX models with anisotropic matter 
satisfying~\eqref{wdomain} and~\eqref{vmdomain},
there exists an open subset of LRS type~IX initial data 
such that the corresponding solutions converge to $\mathrm{R}$.
Because $H_D = 1$ at $\mathrm{R}$ 
we infer from~\eqref{HDinrs} that $H$ is positive
for these solutions for late times; by the 
remark following~\eqref{Heq}
we obtain that $H$ is positive, i.e., these models
are \textit{expanding for all times}.

In the following we analyze the asymptotic behavior
of these forever expanding LRS type~IX solutions in terms of
the metric variables:
For a solution of~\eqref{dynsys}
that converges to the point $\mathrm{R}$
we find
$r\rightarrow 0$, $s\rightarrow 0$, $\Sigma_D \rightarrow -\beta$,
hence $H_D \rightarrow 1$, $N_{1D} \rightarrow 0$ by~\eqref{HDN1Dinrs}.
Furthermore, 
$N_{2D}:= n_2/D \rightarrow \infty$; to see this we first 
note that $s\rightarrow 0$ implies $s \sim N_{1D} N_{2D}^{-1}$ by~\eqref{sinn1n2}.
Then, from $N_{1D} \sim \sqrt{r s}$, see~\eqref{N1Dinrs}, we conclude that $N_{2 D} \sim \sqrt{r/s}$.
Using that $s^{-1} s^\prime|_{\mathrm{R}} = 6 \beta$, cf.~\eqref{sinkprop}, and
$r^{-1} r^\prime|_{\mathrm{R}} = 2 (q + \beta)$, cf.~\eqref{r-1r'},
where
\begin{equation*}
q:= q_D\big|_{\mathrm{R}} =  \textfrac{3}{2} (1- w) \beta^2 + \textfrac{1}{2} ( 1 + 3 w)\:,
\end{equation*}
we finally get $N_{2 D}^{-1} N_{2 D}^\prime|_{\mathrm{R}} = q - 2 \beta > 0$, and the claim follows.
The fact that $N_{2D} = n_2/D \rightarrow \infty$ implies that
$N_2 := n_2/H \rightarrow \infty$ as $\tau\rightarrow \infty$ since $H \simeq D$ in
the limit (because $H_D = H/D = 1$).
This property is completely unproblematic here,
but makes the treatment of the problem extremely difficult
in the standard Hubble-normalized approach.

For large 
$\tau$, the shear variables are
$\sigma_1 = -2\sigma = -2 D \Sigma_D \simeq 2 D \beta$ and
$\sigma_2 = \sigma_3 = \sigma = D \Sigma_D \simeq -D \beta$.
To obtain the metric we 
integrate 
$\partial_t g_{kk} = 2 g_{kk} ( H + \sigma_k )$ 
(which is a consequence of $\partial_t g_{i j} = -2 g_{i l} k^l_{\weg j}$).
In the first step we note that $H = H_D D = D$ so that
the equation reads $\partial_t g_{kk} = 2 g_{kk}  ( D + \sigma_{k} )$, 
i.e.,
\begin{equation}\label{partialgs}
\partial_\tau g_{11} = 2 g_{11}  ( 1 + 2 \beta )\:,\quad
\partial_\tau g_{22} = 2 g_{22}  ( 1 -  \beta )\:. 
\end{equation}
Second, we integrate $D^{-1} D^\prime|_{\mathrm{R}} = -(1+q)$
and~\eqref{newtime} to get $\tau = (1 + q)^{-1} \log(t-t_0) + \tau_0$
with constants $\tau_0$ and $t_0$.
Therefore,~\eqref{partialgs} leads to
an asymptotic behavior of the metric represented by
\begin{equation}\label{metricexp}
g_{11} \propto (t- t_0)^{\frac{2(1+2\beta)}{1+q}} \:,\quad
g_{22} \propto (t- t_0)^{\frac{2(1-\beta)}{1+q}}  
\end{equation}
as $t\rightarrow \infty$.
Note that $\beta$ is negative, but $1 + 2\beta$ is positive, 
which is a consequence of~\eqref{vmdomain}.
An interesting observation is the occurence of partial (directional) 
accelerated expansion. A straightforward calculation reveals
that $(1-\beta)/(1+q) > 1$, hence lengths in the plane
of local rotational symmetry expand at an accelerated rate.
The maximal rate of acceleration is obtained by maximizing
$(1-\beta)/(1+q)$ over the domain depicted
in 
Fig.~\ref{vpmminmax}; 
the maximal value of $(1-\beta)/(1+q) \approx 1.112$
is attained for $w$ close to $-1/3$. 

We conclude by restating the main result: 
The behavior~\eqref{metricexp} is typical, i.e., there exists
an open set of LRS type~IX initial data such that
the associated solutions of the Einstein equations with anisotropic
matter behave as~\eqref{metricexp} as $t\rightarrow \infty$
and thus expand forever.
(These solutions do not behave extraordinarily in other respects;
for instance, there exist forever expanding solutions 
that isotropize toward the singularity.)
In this sense, eternal expansion is as likely as recollapse
in the case of LRS Bianchi type~IX with anisotropic
matter that satisfies the conditions~\eqref{wdomain} 
and~\eqref{vmdomain}, and thus in particular the strong energy condition.

\nextparagraph
\paragraph{Acknowledgements.}

We gratefully acknowledge the hospitality of the
Mittag-Leffler Institute, Sweden.
SC is supported by Ministerio Ciencia e Innovaci\'on, Spain
(Project MTM2008-05271).


\end{document}